\documentclass{kluwer}    
\usepackage{graphicx}
\usepackage{epsfig}

\newdisplay{guess}{Conjecture}

\begin{document}

\begin{article}
\begin{opening}         
\title{M51 revisited: a genetic algorithm approach of its interaction history}

\author{Christian \surname{Theis}\thanks{theis@astrophysik.uni-kiel.de}}
\author{Christian \surname{Spinneker}}

\runningauthor{Ch.\ Theis \& Ch.\ Spinneker}
\runningtitle{M51 revisited}
\institute{Institut f.\ Theoretische Physik und Astrophysik, 
Universit\"at Kiel, 24098 Kiel, Germany}

\begin{abstract}

    Detailed models of observed interacting galaxies suffer from the
extended parameter space. Here, we present results from our code 
{\sc minga} which couples an evolutionary optimization strategy
(a genetic algorithm) with a fast N-body method. {\sc minga} allows
for an automatic search of the optimal region(s) in parameter space
within a few hours to a few days of CPU time on a modern PC by investigating
of the order of $10^5$ models.
We demonstrate its applicability by modelling the 
H{\sc i} intensity and velocity maps of the interacting system M51 and 
NGC 5195.
We get a good fit for the H{\sc i} intensity map and we can reproduce the
counter-rotation feature of the H{\sc i} arm.
Our result corroborates the results of Salo \& Laurikainen (2000)
who favour a model with multiple passages through M51's disk.

\end{abstract}

\end{opening}           


\section{Introduction}

   Interacting galaxies are a rich source for studying many different 
astrophysical phenomena. The perturbation exerted
by a companion can result in a strong tidal response like a 
bridge or tidal tails. On smaller scales the perturbed interstellar
medium might react with induced star formation, maybe even with a 
star burst or the formation of a tidal dwarf galaxy. On larger
scales, the appearance of a galaxy might change substantially, e.g.\
when spiral galaxies are transformed into ellipticals.
Additionally, the evolution of tidal tails is rather sensitive 
to the galactic mass distribution, especially to the dark matter 
profile in the outer regions. Thus, detailed models of interacting galaxies
give not only their dynamical interaction history, but they 
provide us with constraints on star formation timescales or they allow for
measurements of the mass distribution in galaxies.

   The main difficulties for modelling interacting galaxies are the
large parameter space and the substantial CPU-time of a single self-consistent
simulation. Though there has been a tremendous
increase in computational power, we are still completely unable to
cover the whole parameter space by self-consistent N-body simulations. 
E.g.\ covering a
7-dimensional parameter space by 5 values per dimension and assuming 
3 CPU-h per simulation results in about 27 CPU-years. Thus, a first
step in detailed modelling must be an efficient search in parameter space
for interesting regions followed by detailed investigations of these
regions. Genetic algorithms combined with the fast, but approximative
restricted N-body method have proven to be a good tool for such an investigation
(\opencite{wahde98}; \opencite{theis99}). Here, we present first results of 
our modelling of M51. We derived our results directly from 
high resolution H{\sc i} intensity and velocity maps by this proving that 
our code {\sc minga} allows not only for uniqueness checks of preferred model
scenarios, but also for an automatic fit of high-resolution observations.


\subsection*{The M51 encounter in a nutshell}

  M51 is a prototype interacting galaxy which has been modelled many times.
E.g.\ \inlinecite{toomre72} presented a model derived from restricted N-body
simulations giving a good representation of the optical image. Their best
model corresponds to an elliptic (but close to parabolic) orbit with a perigalactic
passage $10^8$ years ago at a distance of 15 kpc\footnote{scaled to a distance 
of 8.4 Mpc to M51.}. They assumed a mass ratio of 1:3 between NGC 5195 and M51. Based
on similar initial conditions, \inlinecite{hernquist90} performed self-consistent 
simulations using a TREE-code. In addition to the previous models, the self-gravity
allowed for the emergence of a central spiral structure whereas the results
in the outer regions are rather similar to the results of the restricted
N-body calculations. 

  The picture became more complicated when H{\sc i} observations
were available: \inlinecite{rots90} found an extended, lopsided 
H{\sc i} tail with a mass of $\sim 5 \cdot 10^8 \cal{M}_\odot$. This tail cannot 
be explained by the earlier models based on optical data. It has a projected length 
of about 80 kpc and a width of 9 kpc. Parts of the H{\sc i} arm are in
counter-rotation with respect to regions north of the center of M51. Additionally, 
a high-speed gas lump was found close to the position of NGC 5195.
Several small but regularly organized H{\sc i} clumps have been detected north 
of NGC 5195. Recently, \inlinecite{salo00} found two regions in parameter space
which give a good match to the intensity maps. Both of them are characterized
by elliptic orbits with similar galactocentric passages. However, they differ
by the orientation of their angular momentum and the number of recent
passages through the disk of M51. Taking the velocity data into account their
models seem to favour multiple disk passages. In another set of models 
-- also using a genetic algorithm -- \inlinecite{wahde01} reproduced the H{\sc i} 
maps with a hyperbolic encounter, but they failed for the kinematics.

\begin{figure}[t]
  \centerline{\hbox{
  \psfig{figure=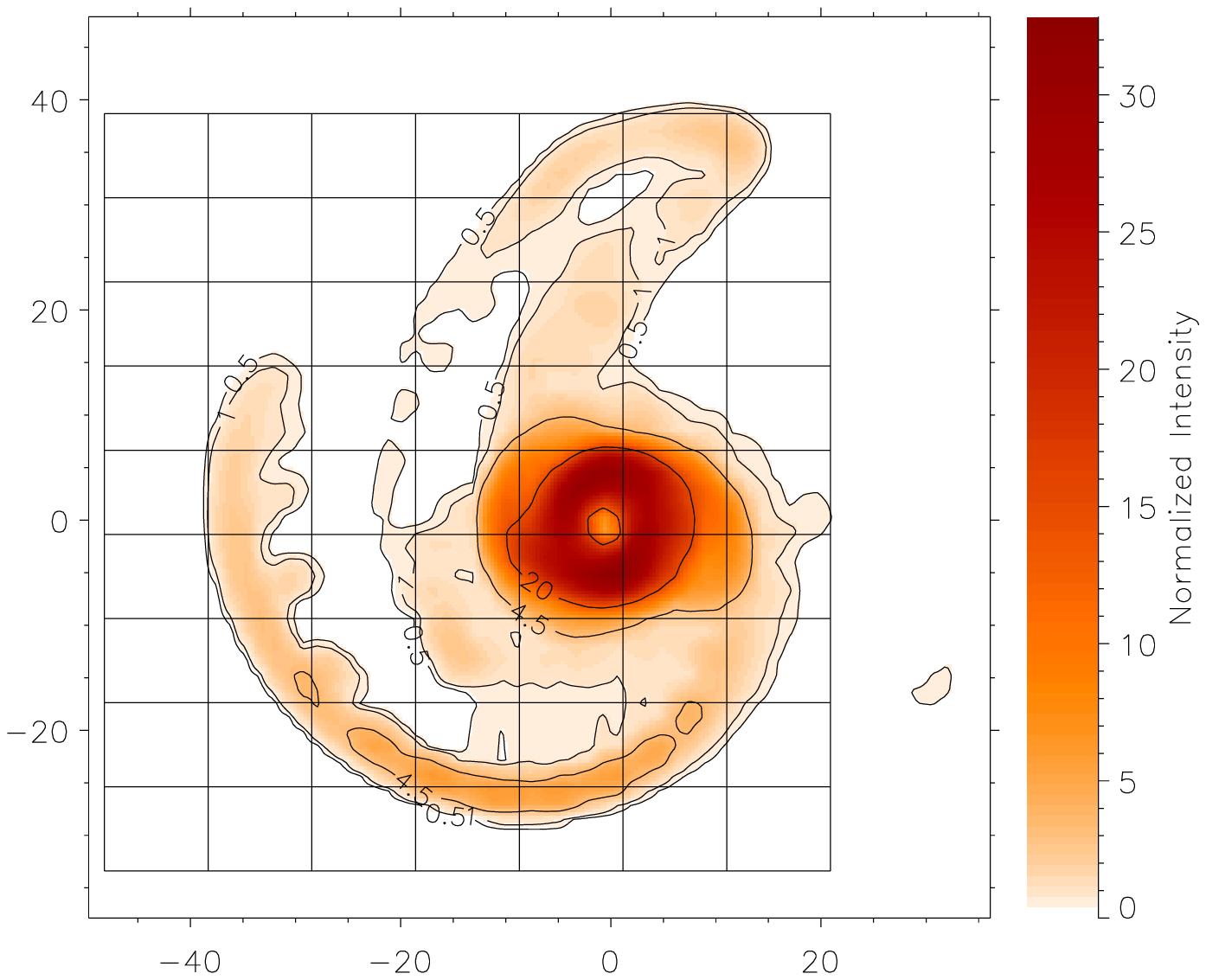,width=7.0cm,angle=0}
  \psfig{figure=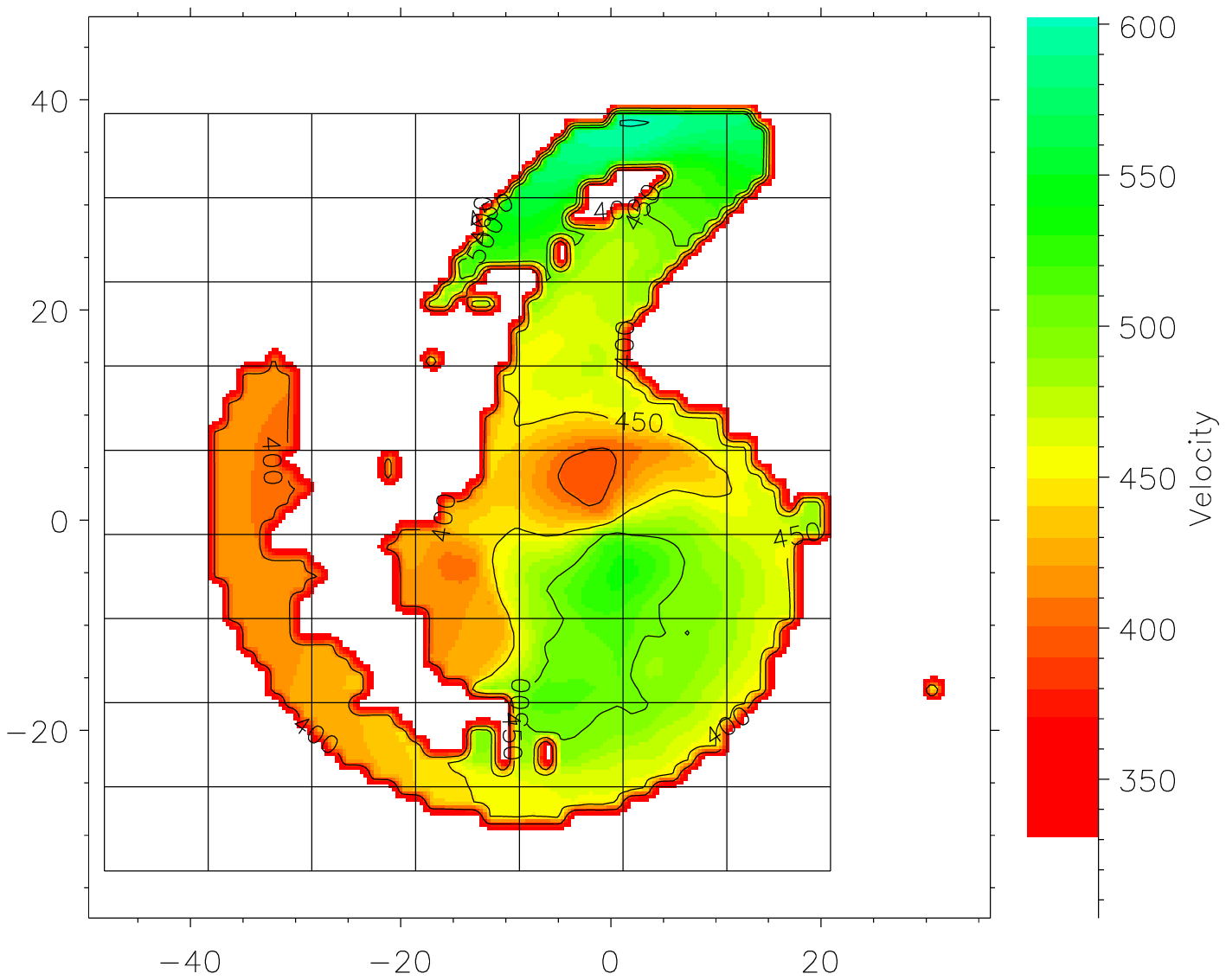,width=7.0cm,angle=0}
  }}
  \caption{Intensity and velocity map derived from the {\sc minga} fit.} 
  \label{theisfig1}
\end{figure}

\section{The {\sc minga} modelling}


  Optimization schemes based on a genetic algorithm (GA) mimic natural evolution 
assuming that the ''population'' improves its adaptation to the ''environmental'' 
constraints during the course of evolution. Details of the implementation of that
concept for modelling interacting galaxies, especially with respect to our code
{\sc minga} can be found e.g.\ in \inlinecite{theis01}. Additionally, we 
implemented a couple of new features in {\sc minga} like masking and emphasizing
parts of the reference images, a direct usage of observational maps (FITS images)
including both intensity and velocity maps,
a consistent treatment of dark halo potentials for the orbit determinations,
different reference map geometries and an extended set of fitness evaluation
functions.

\begin{figure}[t]
  \centerline{\hbox{
  \psfig{figure=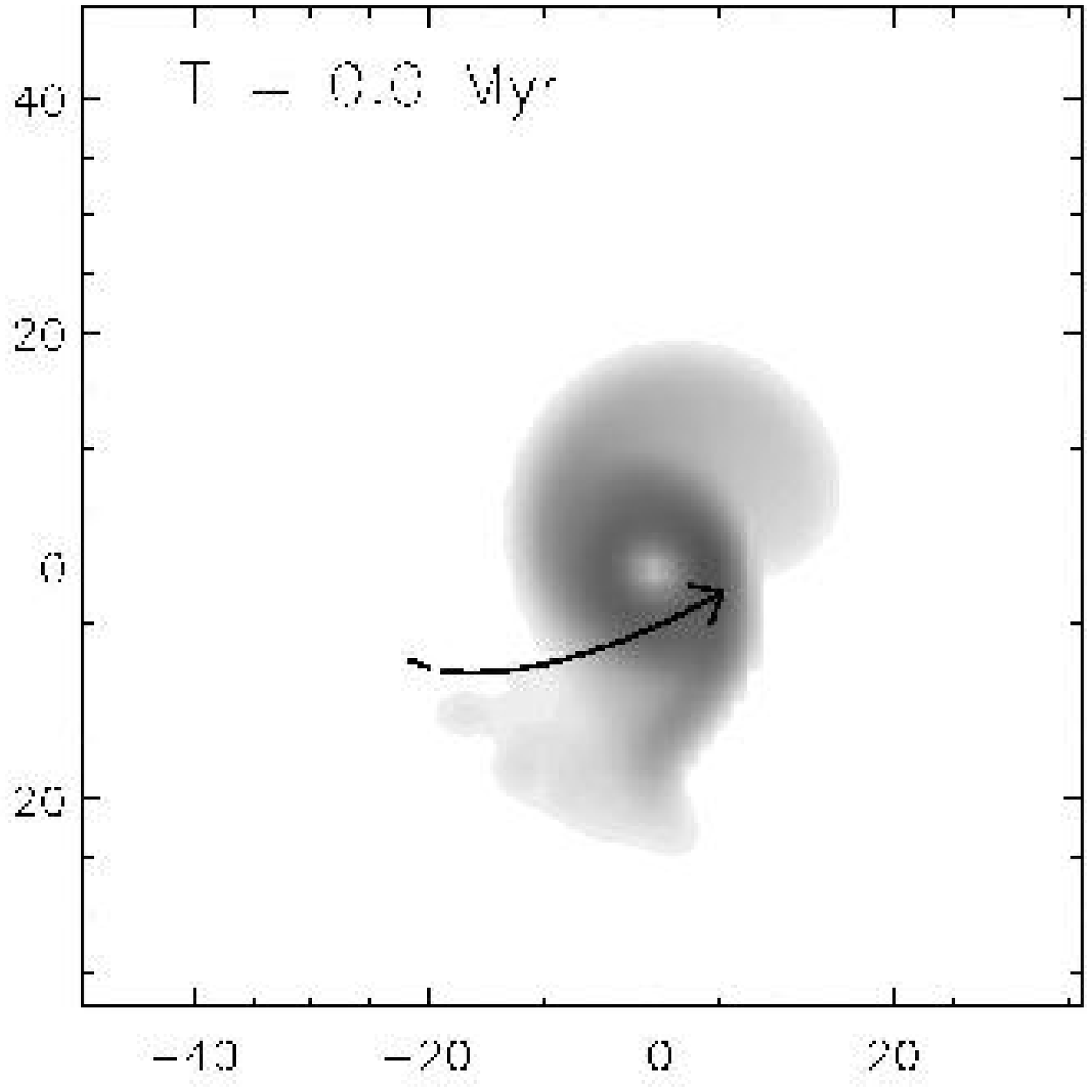,width=3.9cm,angle=0}
  \psfig{figure=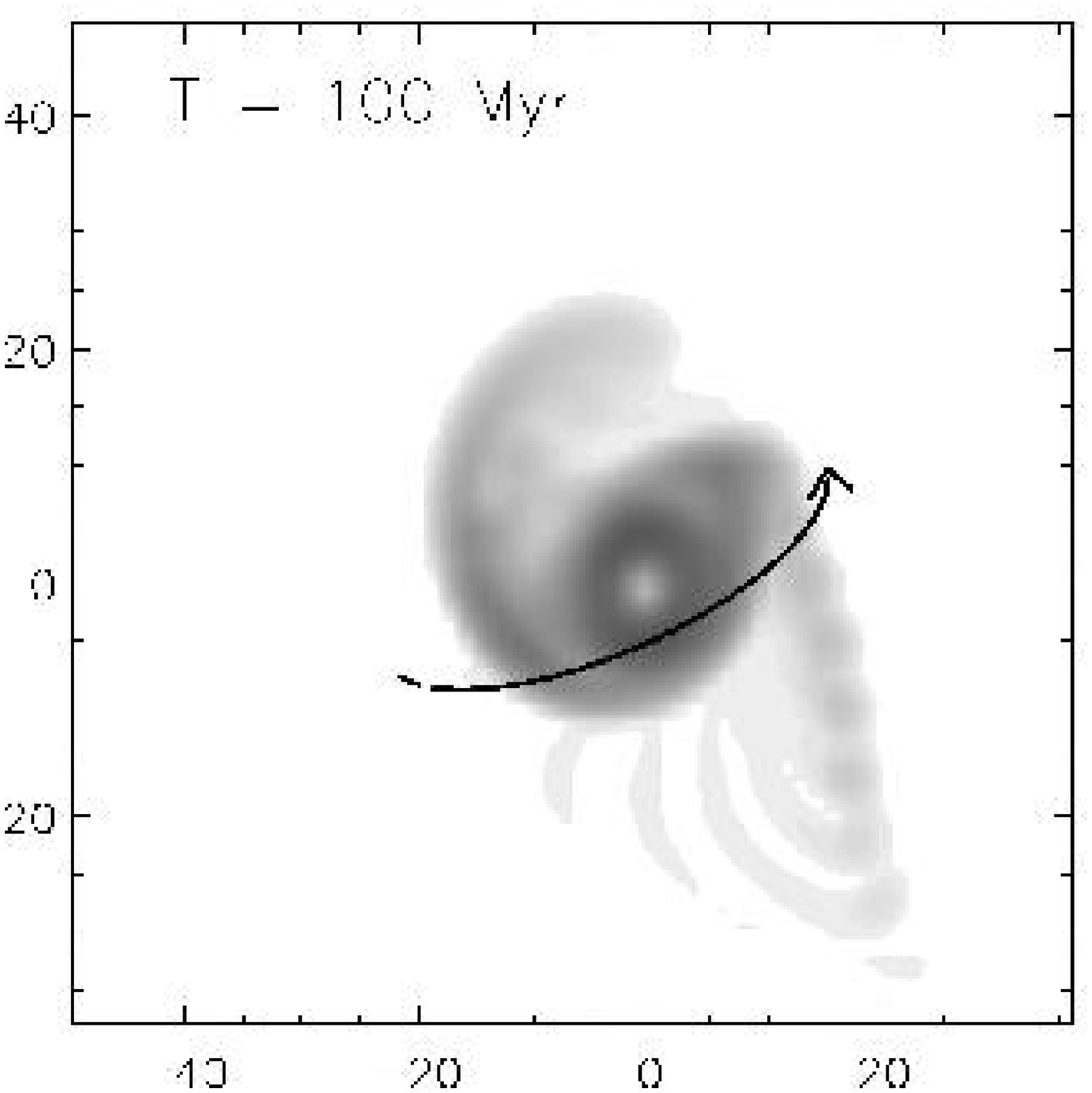,width=3.9cm,angle=0}
  \psfig{figure=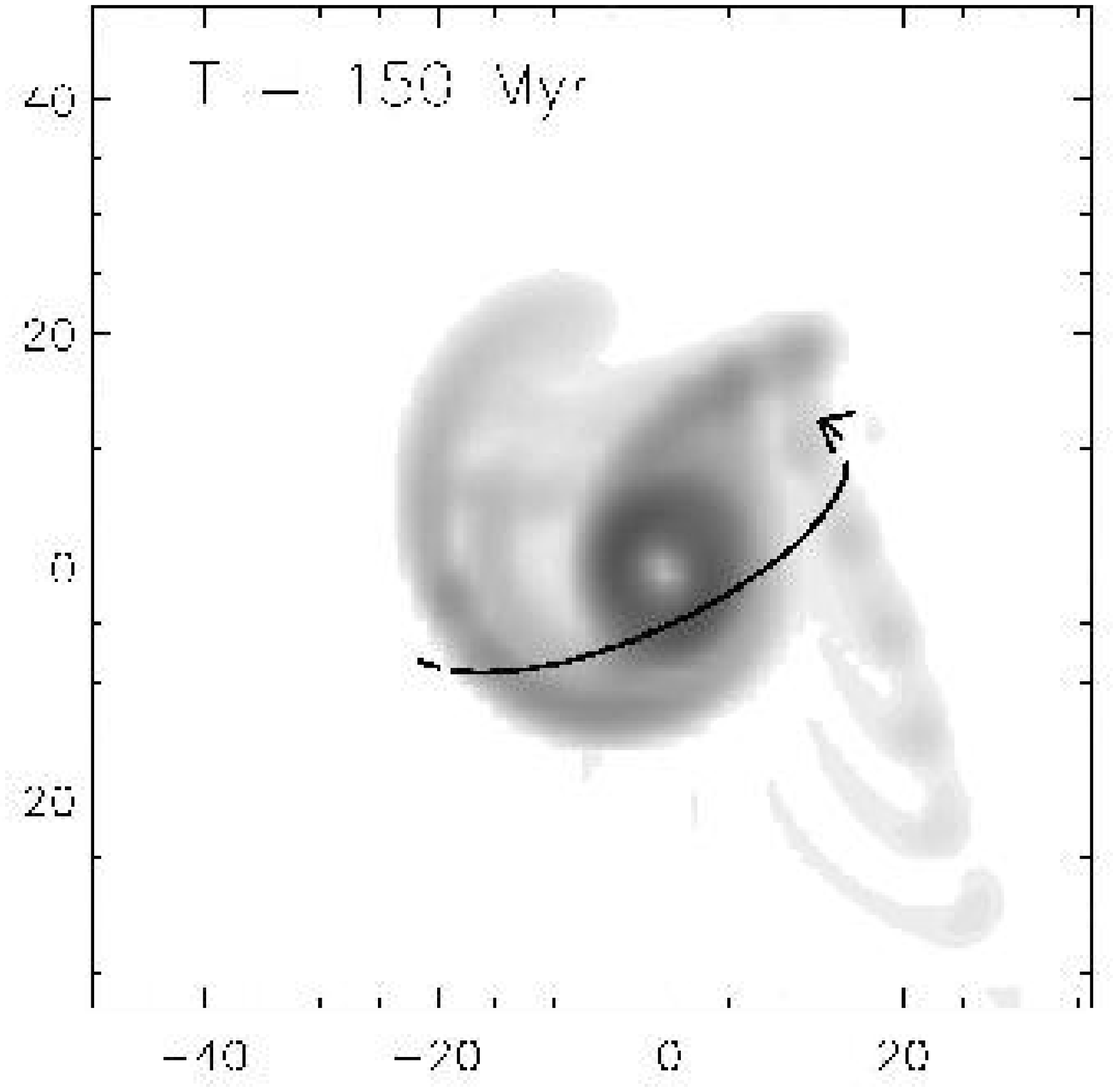,width=3.9cm,angle=0}
  }}
  \centerline{\hbox{
  \psfig{figure=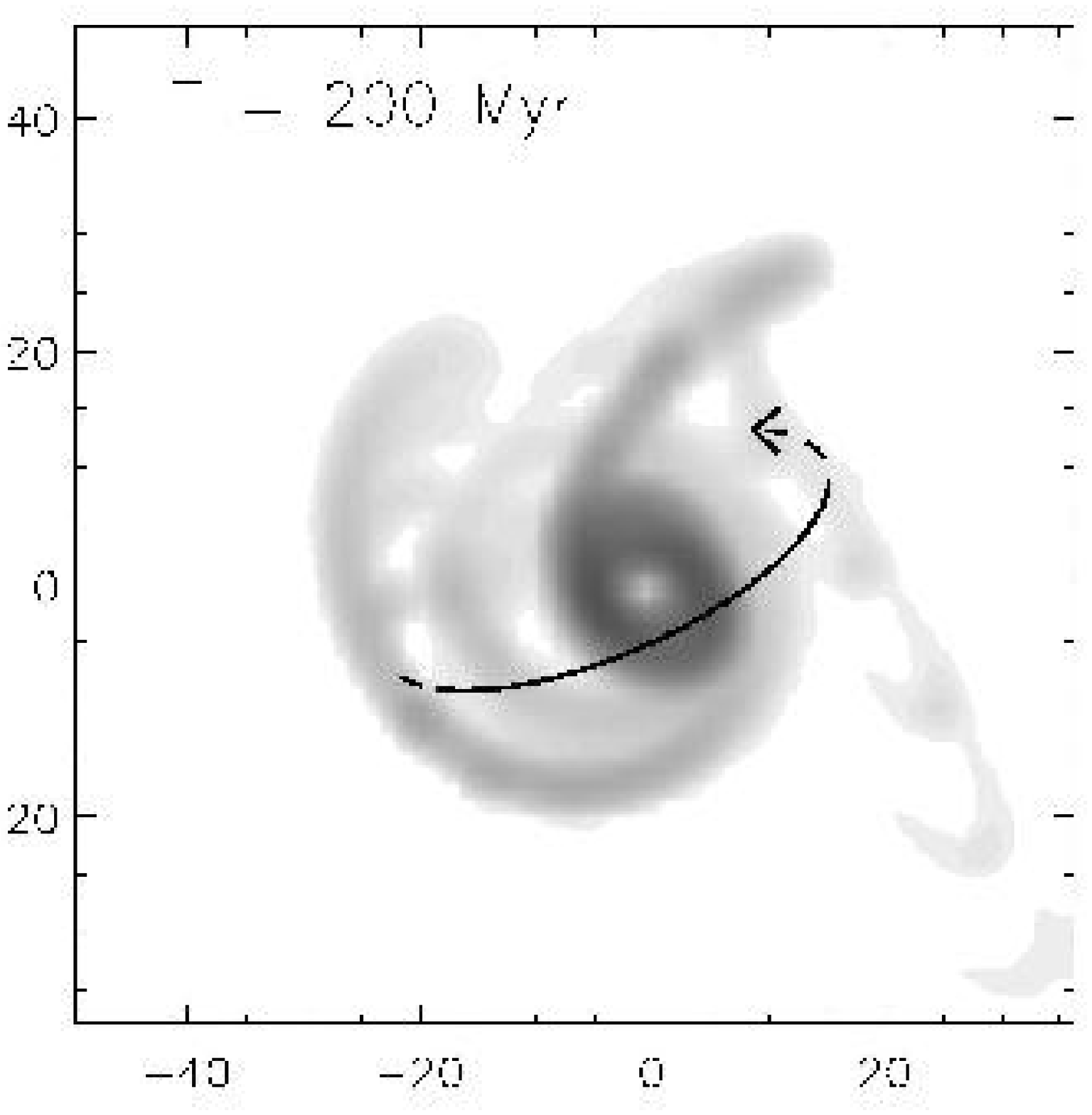,width=3.9cm,angle=0}
  \psfig{figure=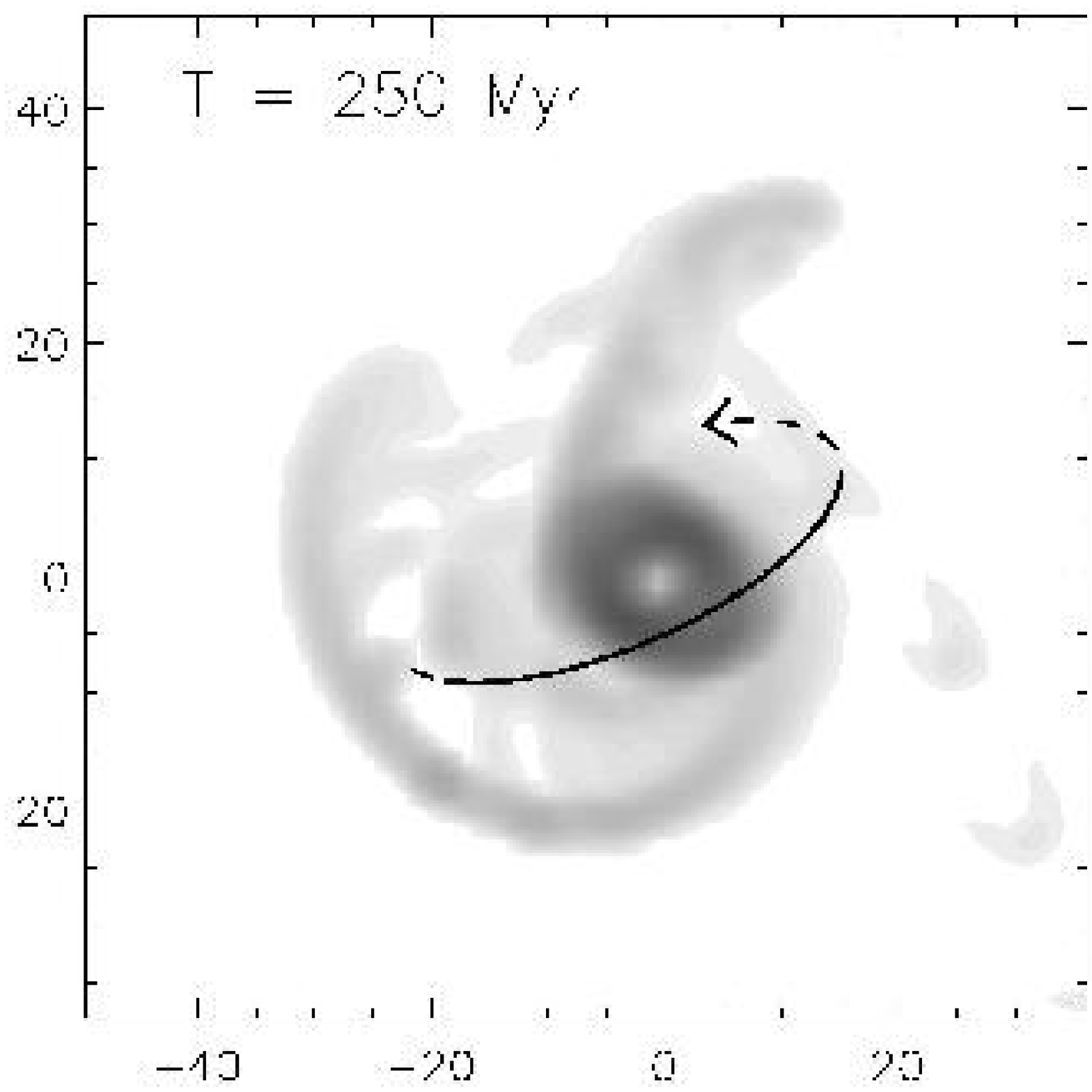,width=3.9cm,angle=0}
  \psfig{figure=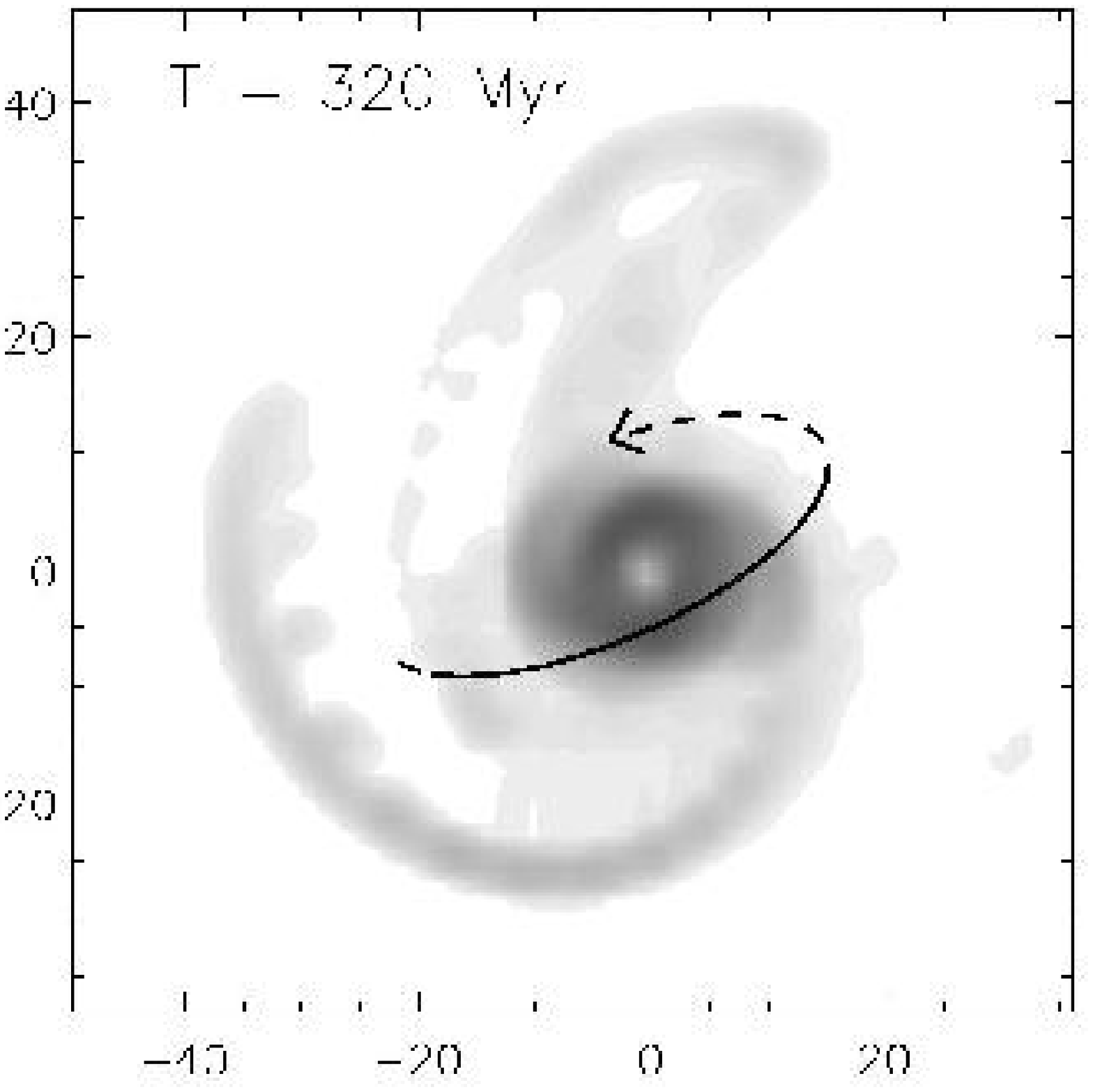,width=3.9cm,angle=0}
  }}
  \caption{Temporal evolution of the H{\sc i} distribution in the M51 system since 
  the last perigalactic passage. Shown is the projection onto the plane of sky.}
  \label{theisfig2}
\end{figure}

    In a first series of {\sc minga} runs we constrained the interaction
by using {\it only} the H{\sc i} intensity map. 
Though we got quickly a nice fit, the best
models were unphysical, because the companion's final position
was in front of M51 clearly violating optical observations. 
In a second series of GA runs we included also the H{\sc i} velocity map for the 
evaluation of the fitness function by calculating a weighted mean of 
the matches to both, 
the intensity and the velocity map. Additionally, we increased the size of the
population to 500 and followed the evolution for 500 generations.
By this, we found solutions qualitatively similar to the
observations (Fig.\ \ref{theisfig1}): the H{\sc i} intensity map is very nicely 
reproduced, including the H{\sc i} clumps north of NGC 5195. Additionally, the 
counter-rotation is recovered. The only feature which is still missing in all
reasonable models is the high velocity lump north of M51. The parameters for our best
model give a mass ratio between NGC 5195 and M51 of about 0.3, a highly elliptic 
orbit and a pericentric distance of about 12 kpc. The evolution since the last 
perigalacticon is shown in Fig.\ \ref{theisfig2}.

\section{Summary}

  By combining intensity and velocity maps our genetic algorithm code
{\sc minga} was able to perform an automatic fit for modelling the interaction 
between M51 and NGC 5195. Our best model corresponds to a highly elliptic orbit
with two recent passages of NGC 5195 through the disk of M51. Thus, our calculations 
corroborate the results of \inlinecite{salo00}.

{\bf Acknowledgements.}
We are grateful to L. Athanassoula and A. Bosma for providing the
H{\sc i} data on M51 and to P. Charbonneau and B. Knapp for making their 
program {\sc pikaia} available. 



\end{article}
\end{document}